\documentstyle[12pt,epsfig]{article}
\title{Quantum Stephani exact cosmological solutions and the selection of time variable}
\author{P. Pedram, S. Jalalzadeh\thanks{Email: s-jalalzadeh@sbu.ac.ir}\,\, and S. S. Gousheh
\\ {\small Department of Physics, Shahid Beheshti University,
Evin, Tehran 19839, Iran}}
\begin{document}
\maketitle \baselineskip 24pt
\begin{abstract}
We study perfect fluid Stephani quantum cosmological model. In the
present work the Schutz's variational formalism which recovers the
notion of time is applied. This gives rise to Wheeler-DeWitt
equation for the scale factor. We use the eigenfunctions in order to
construct wave packets for each case. We study the time-dependent
behavior of the expectation value of the scale factor, using
many-worlds and deBroglie-Bohm interpretations of quantum mechanics.
\end{abstract}

\textit{Pacs}:{ 98.80.Qc, 04.40.Nr, 04.60.Ds;}
\section{Introduction}
In recent years observations show that the expansion of the Universe
is accelerating  in the present epoch \cite{1} contrary to
Friedmann-Robertson-Walker (FRW) cosmological models, with
non-relativistic matter and radiation. Some different physical
scenarios using exotic form of matter have been suggested to resolve
this problem \cite{2,3,4,5,6}. In fact the presence of exotic matter
is not necessary to drive an accelerated expansion. Instead we can
relax the assumption of the homogeneity of space, leaving the
isotropy with respect to one point. The most general class of
non-static, perfect fluid solutions of Einstein's equations that are
conformally flat is known as the ``Stephani Universe''
\cite{1-1,2-1}. This model can be embedded in a five-dimensional
flat pseudo-Euclidean space, which is not expansion-free and has
non-vanishing density \cite{9,1-1,3-3}. In general, it has no
symmetry at all, although its three dimensional spatial sections are
homogeneous and isotropic \cite{12}. The spherically symmetric
Stephani Universes and some of their subcases have been examined in
numerous papers \cite{2-1}. So it may be important to study the
quantum behavior of this models.

The notion of time can be recovered in some cases of quantum
cosmology, for example when gravity is coupled to a perfect fluid
\cite{14-2,R22-4,R22-6}. This kind of systems are often studied as
follows \cite{chap,R22-4,R22-6}. First one uses the Schutz's
formalism for the description of the perfect fluid \cite{11-2,12-2},
second one selects the dynamical variable of perfect fluid as the
reference time. Finally, one uses canonical quantization to obtain
the Wheeler-Dewitt (WD) equation in minisuperspace, which is a
Schr\"odinger-like equation \cite{14-2}. After solving the equation,
one can construct wave packets from the resulting modes. The wave
packets can be used to compute the time-dependent behavior of the
scale factor. If the selected time variable results in a close
correspondence between the expectation value of the scale factor and
the classical prediction (prediction of General Relativity) for long
enough time, the selected time variable can be considered as
acceptable. This approach has been extensively employed in the
literature, indicating in general the suppression of the initial
singularity
\cite{pedram,chap,14-2,R23-10,nivaldo-2,R22-4,R23-13,R22-6}.

We can also study this situation from ontological interpretation of
quantum mechanics \cite{R22-4,bohm1,bohm2,bohm3,bohm4,bohm5}. In
this approach, the problem of time is solved in any situation
(not only in the presence of the matter field). In fact, the ontological
interpretation predicts that the system follows a real trajectory
given by
\begin{equation}\label{dbb}
p_q = S_{,q},
\end{equation}
where $p_q$ is the momentum conjugated to the variable $q$. Here,
$S$ is the phase of the wave function ($\Psi =Re^{iS}$), where $R$
and $S$ are real functions. The equation of motion (\ref{dbb}) is
governed not only by a classical potential $V$ but also by the following quantum
potential $V_Q = \frac{\displaystyle\nabla^2R}{\displaystyle R}$.

In the present paper, we use the formalism of quantum cosmology in
order to quantize Stephani cosmological model in Schutz's formalism
\cite{11-2,12-2} and find WD equation in minisuperspace. In the
Schutz's variational formalism the wave function depends on the
scale factor $a$ and on the canonical variable associated to the
fluid, which  plays the role of time $T$. Here, we describe matter
as a perfect fluid matter. Although, this is semiclassical from the
start, but it has the advantage of defining a variable, connected
with the matter degrees of freedom which can be identified with time
and leads to a well-defined Hilbert space.

\section{The Model}
The action for gravity plus perfect fluid in Schutz's formalism is
written as
\begin{eqnarray}
\label{action} S =\int_Md^4x\sqrt{-g}\, R + 2\int_{\partial
M}d^3x\sqrt{h}\, h_{ab}\, K^{ab}+ \int_Md^4x\sqrt{-g}\, p,
\end{eqnarray}
where $K^{ab}$ is the extrinsic curvature and $h_{ab}$ is the
induced metric over the three-dimensional spatial hypersurface,
which is the boundary $\partial M$ of the four dimensional manifold
$M$ in units where $16\pi G=1$ \cite{7-2}. The last term of
(\ref{action}) represents the matter contribution to the total
action, where $p$ is the pressure. In Schutz's formalism
\cite{11-2,12-2} the fluid's four-velocity is expressed in terms of
five potentials $\epsilon$, $\zeta$, $\xi$, $\theta$ and $\tau$:
\begin{equation}
u_\nu = \frac{1}{\mu}(\epsilon_{,\nu} + \zeta\xi_{,\nu} + \theta
\tau_{,\nu}),
\end{equation}
where $\mu$ is the specific enthalpy, the variable $\tau$ is the
specific entropy, while the potentials $\zeta$ and $\xi$ are
connected with rotation and are absent in models of the FRW type.
The variables $\epsilon$ and $\theta$ have no clear physical
meaning. The four-velocity satisfies the following normalization
condition
\begin{equation}
u^\nu u_\nu = -1.
\end{equation}
The metric in spherically symmetric Stephani Universe
\cite{9,10,1-1,12,2-1,14} has the following form,
\begin{eqnarray}\label{metric}
ds^2 = -\left[F(t)
\frac{a(t)}{V(r,t)}\frac{d}{dt}\left(\frac{V(r,t)}{a(r,t)}\right)
\right]^2dt^2+
\frac{a^2(t)}{V^2(r,t)}\left(dr^2+r^2d\Omega^2\right),
\end{eqnarray}
where the functions $V(r,t)$ and $F(t)$ are defined as
\begin{eqnarray}
  V(r,t) &=& 1+\frac{1}{4}k(t)r^2, \\
  F(t) &=& \frac{a(t)}{\sqrt{C^2(t)a^2(t)-k(t)}}.
\end{eqnarray}
Using the line element (\ref{metric}) and the Einstein's equation,
one can easily show the functions $C(t)$, $k(t)$ and $a(t)$ are
not all independent but are related to each other with the
expressions
\begin{eqnarray}
  \rho(t) &=& \frac{3C^2(t)}{8\pi G},\\
  p(r,t) &=& \frac{1}{8\pi G}[2C(t)\dot C(t)\frac{V(r,t)/a(t)}{(V(r,t)/a(t)\dot)}
  - 3C^2(t)],
\end{eqnarray}
where an overdot denotes a derivative with respect to $t$. Note that
in the spherically symmetric Stephani models with the given
coordinate system, the energy density $\rho(t)$ is uniform, contrary
to the pressure $p(r,t)$, which is not uniform and depends on the
distance from the symmetry center at $r = 0$. This is the reason for
the absence of the barotropic equation of state ($p = p(\rho)$) in
such models. However, if we assume some relations between $\rho(t)$
and $p(r,t)$, this could allow us to eliminate one of the unknown
functions such as $C(t)$. Therefore we are left with two unknown
functions $k(t)$ and $a(t)$. The first one $k(t)$, plays the role of
a spatial curvature, and the second one $a(t)$, is the Stephani
version of the FRW scale factor.

Now, we consider an observer which is placed at the symmetry center
of the spherically symmetric Stephani Universe and our physical
assumptions will concern in the vicinity of $r\approx0$. First, we
assume that locally, matter in the Universe satisfies the barotropic
equation of state
\begin{equation}\label{eqbarotropic}
 p(r\approx0, t) =\alpha \rho (t).
\end{equation}
By substituting the Stephani metric (Eq. (\ref{metric})) in the action (Eq. (\ref{action})) and
Choosing a curvature function $k(t)$ in the form \cite{Godlowski}
\begin{equation}\label{k(t)}
k(t)=\beta a^{\gamma}(t),
\end{equation}
and after some thermodynamical considerations \cite{14-2}, the final reduced effective action near
$r\approx0$, takes the form
\begin{equation}
S = \int dt\biggr[-6\frac{\dot a^2a}{N} +6\beta Na^{1+\gamma} +
N^{-1/\alpha} a^3\frac{\alpha}{(\alpha + 1)^{1/\alpha +
1}}(\dot\epsilon + \theta\dot \tau)^{1/\alpha + 1}\exp\biggr(-
\frac{\tau}{\alpha}\biggl) \biggl].
\end{equation}
The reduced action may be further simplified by canonical methods
\cite{14-2} to the super-Hamiltonian
\begin{equation}
{\cal H} = - \frac{p_a^2}{24a} -6\beta a^{1+\gamma} +\frac{
 p_\epsilon^{\alpha + 1}e^{\tau}}{a^{3\alpha}},
\end{equation}
where $p_a= -12{\dot aa}/{N}$ and $p_\epsilon = -\rho_0 u^0 Na^3$,
$\rho_0$ being the rest mass density of the fluid. Using the
canonical transformations
\begin{eqnarray}
T&=&p_{\tau}e^{-\tau}p_\epsilon^{-(\alpha + 1)},\quad \quad  \quad p_T =
p_\epsilon^{\alpha + 1}e^{\tau}  , \nonumber\\
\bar\epsilon &=& \epsilon - (\alpha + 1)\frac{p_{\tau}}{p_\epsilon},
\quad \quad \quad \bar p_\epsilon = p_\epsilon,
\end{eqnarray}
which are the generalization of the ones used in Ref.~\cite{14-2},
the super-Hamiltonian takes the form
\begin{equation}
{\cal H} = - \frac{p_a^2}{24a} - 6\beta a^{1+\gamma} +
\frac{p_T}{a^{3\alpha}} \,\, ,\label{EqHamiltonian}
\end{equation}
where the momentum $p_T$ is the only remaining canonical variable
associated with matter which appears linearly in the
super-Hamiltonian.

The classical dynamics is governed by the Hamilton equations,
derived from Eq. (\ref{EqHamiltonian}) and Poisson brackets, namely
\begin{equation}
\left\{
\begin{array}{llllll}
\dot{a} =&\{a,N{\cal H}\}=-\frac{\displaystyle Np_{a}}{\displaystyle 12a}\, ,\\
 & \\
\dot{p}_{a} =&\{p_{a},N{\cal H}\}=- \frac{\displaystyle N
p_a^2}{\displaystyle 24a^2}+6N(1+\gamma)\beta a^{\gamma}+
\frac{\displaystyle 3N\alpha\, p_T }{\displaystyle a^{1+3\alpha}} \, ,\\
& \\
\dot{T} =&\{T,N{\cal H}\}=Na^{-3\alpha}\, ,\\
 & \\
\dot{p}_{T} =&\{p_{T},N{\cal H}\}=0\, .\\
& \\
\end{array}
\right. \label{4}
\end{equation}
We also have the constraint equation ${\cal H} = 0$. Choosing the
gauge $N=a^{3\alpha}$, we have the following equations for the
system
\begin{eqnarray}\label{eqm1}
T&=&t,\\\label{eqm2} \ddot{a}&=&(3\alpha-\frac{1}{2})\frac{\dot
a^2}{a}-\frac{1}{2}(1+\gamma)\beta
a^{6\alpha+\gamma-1}-\frac{\alpha}{4}a^{3\alpha-2}
p_T,\\\label{eqm3} 0&=&-\frac{6\dot a^2}{a^{6\alpha-1}}-6\beta
a^{\gamma+1}+\frac{p_T}{a^{3\alpha}}.
\end{eqnarray}
Note that the classical equations for $\gamma=+1$ case in
Ref.~\cite{Stelmach}, are corresponding with choosing the gauge
$N=1$. In this gauge the constraint equation ${\cal H} = 0$ reduces
to
\begin{eqnarray}
-6a\dot a^2-6\beta a^{\gamma+1}+a^{-3\alpha}p_T=0,
\end{eqnarray}
or
\begin{eqnarray}
\left(\frac{da(t)}{dt}\right)^2+\beta
a(t)=\frac{p_T}{6a^{3\alpha+1}(t)}.
\end{eqnarray}
Imposing the standard quantization conditions on the canonical
variables ($p_a \rightarrow -i\frac{\displaystyle
\partial}{\displaystyle\partial a}$,  $p_T \rightarrow
-i\frac{\displaystyle\partial}{\displaystyle\partial T}$) and
demanding that the super-Hamiltonian operator (\ref{EqHamiltonian})
annihilate the wave function, we are led to the following WD
equation in minisuperspace ($\hbar =1$)
\begin{equation}
\label{sle} \frac{\partial^2\Psi}{\partial a^2} - 144\beta
a^{2+\gamma}\Psi - i24a^{1 - 3\alpha}\frac{\partial\Psi}{\partial t}
= 0.
\end{equation}
According to the equation (\ref{eqm1}) $T=t$ can be associated with
the time coordinate \cite{lemos1999,lemos1998}. Equation (\ref{sle})
takes the form of a Schr\"odinger equation $i\partial\Psi/\partial t
= {\hat H} \Psi$. As discussed in \cite{nivaldo-2,lemos1998}, in
order for the Hamiltonian operator ${\hat H}$ to be self-adjoint the
inner product of any two wave functions $\Phi$ and $\Psi$ must take
the form
\begin{equation}\label{inner}
(\Phi,\Psi) = \int_0^\infty a^{1 - 3\alpha}\Phi^*\Psi da.
\end{equation}
On the other hand, the wave functions should satisfy the following
boundary conditions \cite{lemos1998,18-2}
\begin{equation}
\label{boundary} \Psi(0,t) = 0 \quad \mbox{or} \quad
\frac{\partial\Psi (a,t)}{\partial a}\bigg\vert_{a = 0} = 0.
\end{equation}
The WD equation (\ref{sle}) can  be solved by separation of
variables as follows,
\begin{equation}
\Psi(a,t) = e^{iEt}\psi(a), \label{11}
\end{equation}
where the scale factor dependent part of the wave function
($\psi(a)$) satisfies
\begin{equation}
\label{sle2} -\psi'' + 144 \beta a^{2+\gamma}\psi =24Ea^{1 -
3\alpha}\psi,
\end{equation}
and the prime denotes derivative with respect to $a$.

An interesting feature of the Stephani model is that the spatial
curvature is time-dependent. The recent observational data show that
our Universe is spatially flat. Moreover, negative powers in
equation (\ref{k(t)}) lead to the spatially flat Universe in the
present epoch. In the next section we solve a general class of
exactly solvable models which $\gamma=-(1+3\alpha)$ and find the
corresponding eigenfunctions. Then we construct the wave packets by
appropriate superimposing of these eigenfunctions and compute the
expectation value of the scale factor versus time. After solving the
classical equations exactly, we compare the classical and quantum
solutions and show that these solutions are asymptotically the same.
Moreover, we study the case using de-Broglie Bohm interpretation of
quantum mechanics and find the corresponding bohmian trajectories
which are different from  the classical case for small times due to
the effect of the quantum potential.

\section{Quantum cosmological models with $\gamma=-(1+3\alpha)$}
In this Section, we consider the particular relation between
$\gamma$ and $\alpha$ as $\gamma=-(1+3\alpha)$. In this case the WD
equation (\ref{sle2}) takes the form
\begin{equation}
\psi'' + 24(E-6\beta)a^{1 - 3\alpha}\psi =0,
\end{equation}
and hence equation (\ref{11}) has the following general
time-dependent solutions under the form of Bessel functions
\begin{equation}
\Psi_E(a,t) = e^{i(E-6\beta)t}\sqrt{a}\biggr[c_1J_{\frac{1}{3(1 -
\alpha)}}\biggr(\frac{\sqrt{96(E-6\beta)}}{3(1 -
\alpha)}a^{\frac{3(1 - \alpha)}{2}}\biggl) + c_2Y_{\frac{1}{3(1 -
\alpha)}}\biggr(\frac{\sqrt{96(E-6\beta)}}{3(1 -
\alpha)}a^{\frac{3(1 - \alpha)}{2}}\biggl)\biggl].
\end{equation}
Wave packets can be constructed by superimposing these solutions to
obtain physically allowed wave functions. The general structure of
these wave packets are
\begin{equation}
\Psi(a,t) = \int_{6\beta}^\infty A(E)\Psi_{E}(a,t)dE.
\end{equation}
We choose $c_2 = 0$ to satisfy the first boundary condition of
Eq.~(\ref{boundary}). Defining $r = \frac{\sqrt{96(E-6\beta)}}{3(1 -
\alpha)}$, simple analytical expressions for the wave packet can be
found by choosing $A(E)$ to be a quasi-gaussian function
\begin{equation}
\Psi(a,t) = \sqrt{a}\int_0^\infty r^{\nu + 1}e^{-\kappa r^2 +
i(\frac{3}{32}r^{2} (1 - \alpha)^2+6\beta)t}J_\nu(ra^\frac{3(1 -
\alpha)}{2})dr,
\end{equation}
where $\nu = \frac{1}{3(1 - \alpha)}$ and $\kappa$ is an arbitrary
positive constant. The above integral is known \cite{gradshteyn},
and the wave packet takes the form
\begin{equation}\label{wavepacket}
\Psi(a,t) = a\frac{e^{-\frac{a^{3(1 - \alpha)}}{4B}+6i\beta
t}}{(-2B)^{\frac{4 - 3\alpha}{3(1 - \alpha)}}},
\end{equation}
where $B = \kappa - i\frac{3}{32}(1 - \alpha)^2t$. Now, we can
verify what these quantum models predict for the behavior of the
scale factor of the Universe. By adopting the many-worlds
interpretation \cite{many}, and with regards to the inner product
relation (\ref{inner}), the expectation value of the scale factor
\begin{equation}\label{sum}
\langle a\rangle_t = \frac{\int_0^\infty a^{1 -
3\alpha}\Psi(a,t)^*a\Psi(a,t)da} {\int_0^\infty a^{1 -
3\alpha}\Psi(a,t)^*\Psi(a,t)da},
\end{equation}
is easily computed, leading to
\begin{equation}\label{scale-qm}
\langle a\rangle_t=\frac{\Gamma \left(\frac{3 \alpha -5}{3 (\alpha
-1)}\right)}{\Gamma \left(\frac{3 \alpha -4}{3
   (\alpha -1)}\right)}\biggr[\frac{\frac{18(1 -
\alpha)^4}{(32)^2}t^2 +2 \kappa^2}{\kappa}\biggl]^\frac{1}{3(1 -
\alpha)}.
\end{equation}
Now, we can calculate the dispersion of the wave packets
\begin{equation}
(\Delta a)_t^2=\langle a^2\rangle_t-\langle a\rangle_t^2,
\end{equation}
using (\ref{wavepacket},\ref{sum}), we have
\begin{equation}
(\Delta a)_t^2=\frac{3 \pi  \Gamma \left(\frac{1}{1-\alpha
}\right)-16^{\frac{1}{3-3 \alpha }} \Gamma
   \left(\frac{1}{3-3 \alpha }\right) \Gamma \left(\frac{5-3 \alpha }{6-6 \alpha }\right)^2}{\pi
   \Gamma \left(\frac{1}{3-3 \alpha }\right)}\biggr[\frac{\frac{18(1 -
\alpha)^4}{(32)^2}t^2 +2 \kappa^2}{\kappa}\biggl]^\frac{2}{3(1 -
\alpha)}.
\end{equation}
This shows the dispersion of the wave packets through the time with
the minimum at $t=0$. This is similar to the free particle case,
where the the wave packets disperse more rapidly for more localized
initial states.

The important feature of this model is the avoidance of the
singularity. Equation (\ref{scale-qm}) shows that the expectation
value of the scale factor never vanishes for all time. On the other
hand, at the quantum level, since the probability density of finding
the scale factor at $a=0$ (with regards to the inner product
relation (\ref{inner}) and the behavior of Bessel functions for
small values of the argument) is zero in all times
($\lim_{a\rightarrow0}a^{1-3\alpha}|\Psi(a,t)|^2=0$), we have a
indication that these models may not have singularities at the
quantum level.

\begin{figure}
  \centering
\begin{tabular}{ccc}
\includegraphics[width=7cm]{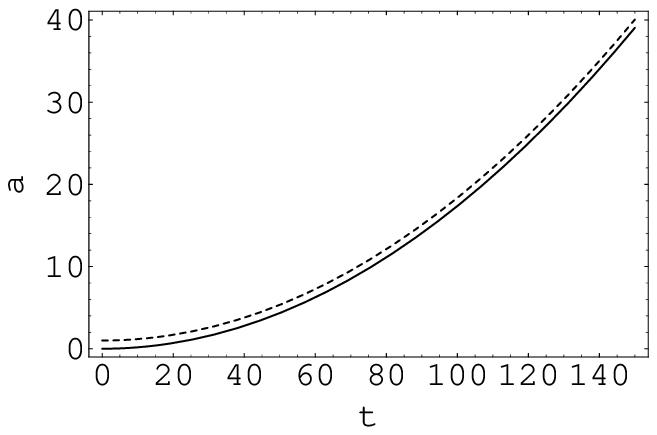}
 &\hspace{1.cm}&
\includegraphics[width=7cm]{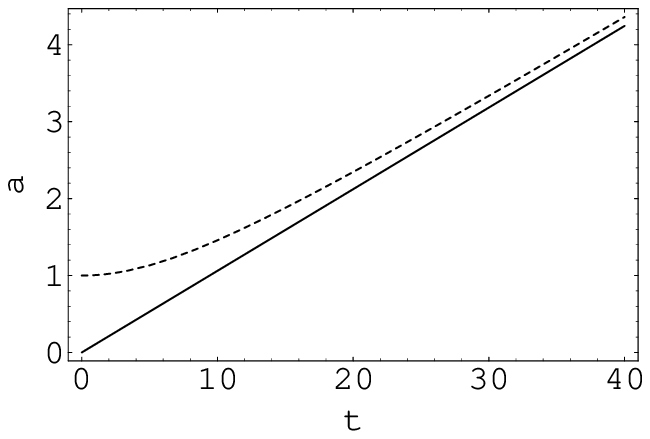}
\end{tabular}
\begin{tabular}{ccc}
\includegraphics[width=7cm]{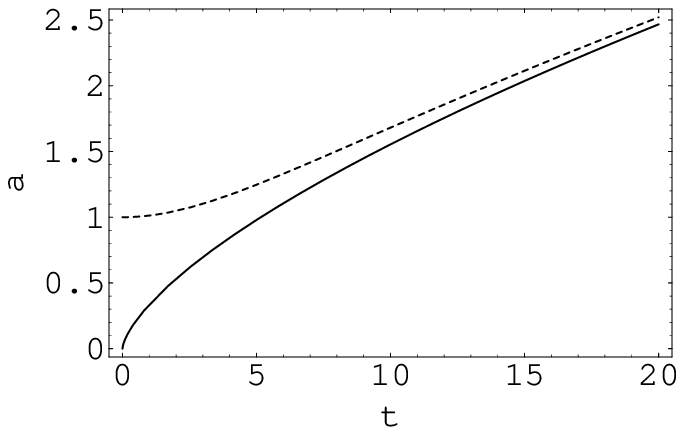}
 &\hspace{1.cm}&
\includegraphics[width=7cm]{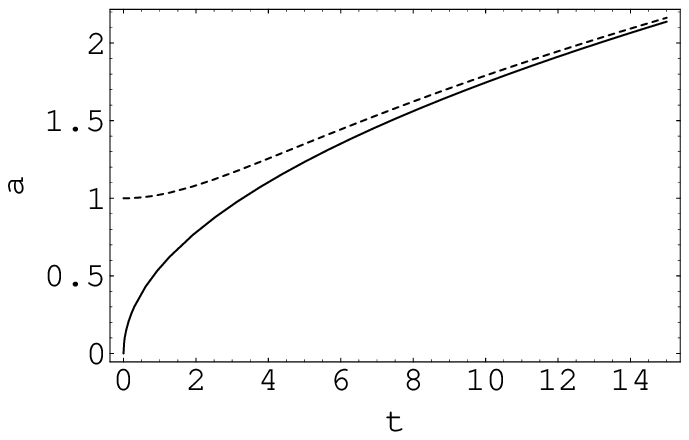}
\end{tabular}
\begin{tabular}{ccc}
\includegraphics[width=7cm]{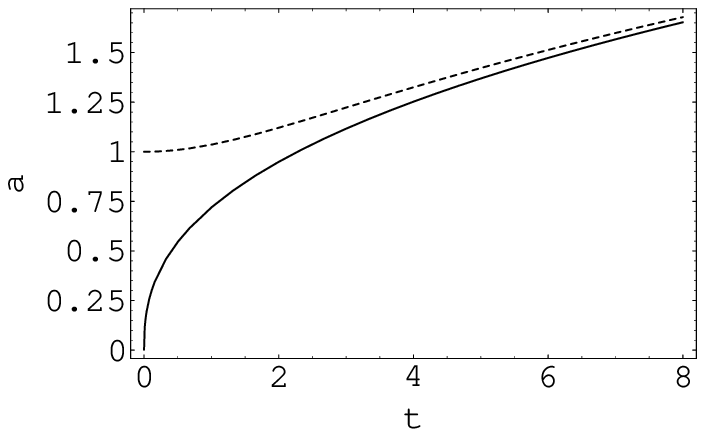}
 &\hspace{1.cm}&
\includegraphics[width=7cm]{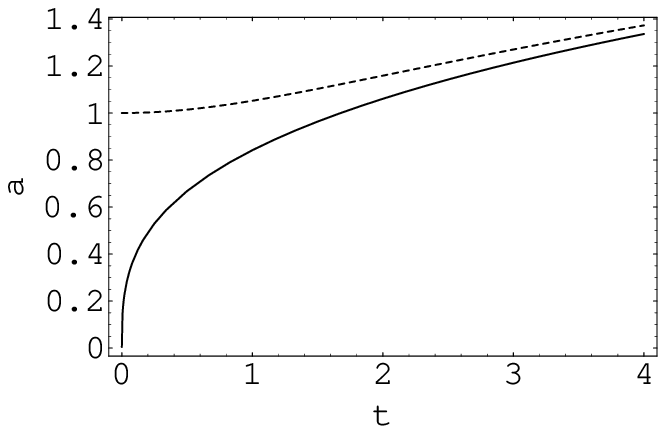}
\end{tabular}
  \caption{Classical scale factor and the expectation value of the scale factor versus time for
  $\alpha=2/3$ (up, left), $\alpha=1/3$ (up, right), $\alpha=0$ (middle, left), $\alpha=-1/3$ (middle, right),
  $\alpha=-2/3$ (down, left), and $\alpha=-1$ (down, right).}\label{fig1}
\end{figure}

In classical case, by eliminating the $p_T$ variable in the
equations of motions (\ref{eqm2},\ref{eqm3}), the resulting
equation in case $\gamma=-(1+3\alpha)$ takes the following simple
form
\begin{equation}
\ddot{a}=(3\alpha-1)\frac{\dot{a}^2}{2a},
\end{equation}
which has the exact solution as
\begin{equation}\label{scale-cl}
a(t)=a_0 t^{\frac{2}{3(1 - \alpha)}},
\end{equation}
where
\begin{equation}
a_0=\frac{\Gamma \left(\frac{3 \alpha -5}{3 (\alpha
-1)}\right)}{\Gamma \left(\frac{3 \alpha -4}{3
   (\alpha -1)}\right)}\biggr[\frac{18(1 -
\alpha)^4}{(32)^2\kappa}\biggl]^\frac{1}{3(1 - \alpha)}.
\end{equation}
Figure \ref{fig1} shows the behavior of the classical scale factor
(\ref{scale-cl}) and quantum mechanical expectation value of the
scale factor (\ref{scale-cl}) versus time for various cases.

It is known that the results obtained by using the many-worlds
interpretation agree with those that can be obtained using the
ontological interpretation of quantum mechanics
\cite{bohm1,bohm2,bohm3,bohm4,bohm5}. In Bohmian interpretation the
wave function is written as
\begin{equation}
\Psi = R\, e^{iS}
\end{equation}
where $R$ and $S$ are real functions. Inserting this expression in
the WD  equation (\ref{sle}) (for $\gamma=-(1+3\alpha)$), we have
\begin{eqnarray}
\label{hje}
\frac{\partial S}{\partial t} - \frac{1}{24a^{1- 3\alpha}}\biggr(\frac{\partial S}{\partial a}\biggl)^2 -6\, \beta+ \, Q &=& 0 ,\\
\frac{\partial R}{\partial t} - \frac{1}{12a^{1 -
3\alpha}}\frac{\partial R}{\partial a}\frac{\partial S}{\partial a}
- \frac{1}{24a^{1 - 3\alpha}}R \frac{\partial^2S}{\partial a^2} &=&
0.
\end{eqnarray}
Here, $Q =
\frac{1}{24a^{1-3\alpha}}\frac{1}{R}\frac{\partial^2R}{\partial
a^2}$ is the quantum potential which modifies the Hamilton-Jacobi
equation. In fact, the deviation from the classical trajectory
happens whenever the quantum potential is more important than the
classical potential. The real functions $R(a,t)$ and $S(a,t)$ can be
obtained from the wave function (\ref{wavepacket}) as
\begin{eqnarray}
R &=& \biggr[4\kappa^2 + \biggr(\frac{3}{16}\biggl)^2(1 -
\alpha)^4t^2\biggl]^{-\frac{4 - 3\alpha}{6(1 - \alpha)}}\, a\,
\exp\biggr\{-\frac{\kappa a^{3(1 - \alpha)}}{4\biggr[\kappa^2 + \biggr(\frac{3}{32}\biggl)^2(1 - \alpha)^4t^2\biggl]}\biggl\}, \\
S &=& - \frac{3}{128}\frac{(1 - \alpha)^2a^{3(1 - \alpha)}t}{
\biggr[\kappa^2 + \biggr(\frac{3}{32}\biggl)^2(1 -
\alpha)^4t^2\biggl]} + \frac{(4 - 3\alpha)}{3(1 -
\alpha)}\arctan\biggr[\frac{3}{32}\frac{(1 -
\alpha)^2t}{\kappa}\biggl]+6\beta t.
\end{eqnarray}
In the Bohmian interpretation, the behavior of the scale factor is
governed by the following equation
\begin{equation}
p_a = \frac{\partial S}{\partial a}.
\end{equation}
On the other hand, from the definition of $p_a$ (\ref{4}) and for $N
= a^{3\alpha}$ we have $p_a=-12a^{1-3\alpha}\dot{a}$. Therefore,
Bohmian trajectory becomes
\begin{equation}
512\frac{\dot a}{a} = 3(1 - \alpha)^3\frac{t}{\biggr[\kappa^2 +
\biggr(\frac{3}{32}\biggl)^2(1 - \alpha)^4t^2\biggl]},
\end{equation}
which the integration yields
\begin{equation}
\label{bt} a(t) = a_0 \biggr[\kappa^2 +
\biggr(\frac{3}{32}\biggl)^2(1 - \alpha)^4t^2\biggl]^\frac{1}{3(1 -
\alpha)},
\end{equation}
where $a_0$ is the constant of the integration. This completely
coincides with the computation of the expectation value of the scale
factor. Now we can find the quantum potential
\begin{eqnarray}
Q(a,t) = - \frac{\kappa}{32}\frac{1 - \alpha}{\biggr[\kappa^2 +
\biggr(\frac{3}{32}\biggl)^2(1 - \alpha)^4t^2\biggl]^2}
\biggr\{3\kappa(1 - \alpha)a^{3(1 - \alpha)} - (4 -
3\alpha)\biggr[\kappa^2 + \biggr(\frac{3}{32}\biggl)^2(1 -
\alpha)^4t^2\biggl]\biggl\}.
\end{eqnarray}
Using the relation between the scale factor and time (\ref{bt}) the
quantum potential can be written in terms of the scale factor as
\begin{equation}
Q(a) = \kappa\frac{1 - \alpha}{32}a_0^{3(1 - \alpha)}\frac{(4 -
3\alpha) - 3\kappa(1 - \alpha)a_0^{3(1 - \alpha)}}{a^{3(1 -
\alpha)}}.
\end{equation}
It is obvious that the quantum effects are negligible for large
values of the scale factor and are important for small values of the
scale factor. Therefore, asymptotically we have the classical
behavior.

In the next section for completeness we consider briefly some interesting
and exactly solvable cases which have bound state solutions.

\section{Bound state solutions}
In this Section, we study four different cases of $\gamma$ and
$\alpha$. We find the exact discrete energy spectrum and
corresponding eigenfunctions.

\begin{figure}
\centering
  \includegraphics[width=8cm]{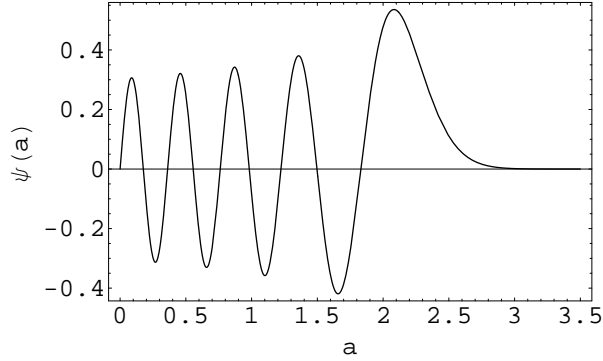}\\
  \caption{Plot of the wave function ($\psi(a)$) for $\beta=1$ and $n=8$, showing the oscillatory behavior for the small values of the scale factor
  and exponential damping for the large values of the scale factor.}\label{fig2}
\end{figure}

For $\gamma=-1$ and $\alpha=1/3$ (radiation), the WD equation
(\ref{sle2}) reduces to
\begin{equation}
\label{radiation-1} -\psi'' + 144 \beta a\psi =24E\psi,
\end{equation}
which can be rewritten as
\begin{equation}
\psi'' -144 \beta\left(a- \frac{\displaystyle E}{\displaystyle 6\beta}\right)\psi =0,
\end{equation}
by taking $x= a-\frac{\displaystyle E}{\displaystyle 6\beta}$ we have
\begin{equation}
\frac{d^2}{dx^2}\psi(x) - 144 \beta x\psi(x) =0,
\end{equation}
which is the Airy's differential equation \cite{Magnus}. This equation has two solutions as
$\mbox{Ai}\left[(144 \beta)^{1/3}x\right]$ and $\mbox{Bi}\left[(144 \beta)^{1/3}x\right]$. First
one is exponentially decreasing function of $x$ and the second one grows exponentially and is
physically unacceptable. Therefore, the solution is
\begin{equation}
\psi(a)= \mbox{Ai}\left[(144 \beta)^{1/3}\left(a-\frac{\displaystyle E}{\displaystyle
6\beta}\right)\right].
\end{equation}
We choose the first boundary condition (\ref{boundary}), which leads to
\begin{equation}
\mbox{Ai}\left[{ -E}{ \left(\sqrt{\frac{3}{2}}\beta\right)^{-2/3}}\right]=0.
\end{equation}
Airy's function $\mbox{Ai}(x)$ has infinitely many negative zeros
$z_n = -a_n$, where $a_n>0$, therefore, the energy levels quantize
and take the values
\begin{equation}
E_n = \left(\sqrt{\frac{3}{2}}\beta\right)^{2/3}\,a_n.
\end{equation}
The time-dependent eigenfunctions take the form
\begin{equation}\label{radiation-final-1}
\Psi_n(a,t)=e^{iE_n  t}\mbox{Ai}\left[(144
\beta)^{1/3}\left(a-\frac{\displaystyle E_n}{\displaystyle
6\beta}\right)\right].
\end{equation}
It is important to note that Airy's function $ \mbox{Ai}(x)$ has an
oscillatory behavior for $x<0$ ($a<\frac{\displaystyle
E_n}{\displaystyle 6\beta}$) whiles for $x>0$
($a>\frac{\displaystyle E_n}{\displaystyle 6\beta}$) decreases
monotonically and is an exponentially damped function for large $x$
(Fig. \ref{fig2}). Therefore, the solutions
(\ref{radiation-final-1}) show a classical behavior for small $a$
and a quantum behavior for large $a$. This is contrary to usually
expected results in the previous Section. In fact detecting quantum
gravitational effects in large Universes is noticeable which has
been also observed in FRW and Kaluza-Klein models
\cite{lemos1999,Coliteste}.

In $\gamma=-1$ and $\alpha=-1/3$ (cosmic strings) case, the WD
equation (\ref{sle2}) reduces to
\begin{equation}
\label{cosmic-1} -\psi'' + 144 \beta a\psi =24Ea^{2}\psi.
\end{equation}
For $E<0$ the above equation can be written as
\begin{equation}
-\psi'' +24 |E|\left[\left(a-\frac{3\beta}{E}\right)^2-\left(\frac{3\beta}{E}\right)^2\right]\psi =0,
\end{equation}
by taking $x=a- \frac{\displaystyle 3\beta}{\displaystyle E}$ we have
\begin{equation}
-\frac{d^2}{dx^2}\psi(x) + 24 |E|x^2\psi(x)
=\frac{216\beta^2}{|E|}\psi(x).
\end{equation}
This equation is identical to the time-independent Schr\"odinger equation for a harmonic oscillator with unit mass and energy $\lambda$:
\begin{equation}
-\frac{d^{2}\psi(x)}{dx^{2}}+\omega^{2}x^{2}\psi(x) =2\lambda \psi(x),
\end{equation}
where $2\lambda = \frac{\displaystyle 216\beta^2}{\displaystyle |E|}$ and $\omega^2=24 |E|$. Therefore, the allowed values of $\lambda$ are
$\omega(n+1/2)$ and the possible values of $E$ are
\begin{equation}
E_{n}=-\left(\frac{113\beta^2}{\sqrt{24}(n+\frac{1}{2})}\right)^{\frac{2}{3}}\,\, , \mbox{\hspace{0.8cm}} n=0,1,2,...\quad .
\end{equation}
therefore, the  stationary solutions are
\begin{equation}\label{cosmic-final-1}
{\Psi}_{n}(a,t)=e^{iE_{n}t}{\varphi}_{n}\left(a- \frac{\displaystyle
3\beta}{\displaystyle E_n}\right),
\end{equation}
\begin{equation}
{\varphi}_{n}(x)=H_n\left((24|E|)^{\frac{1}{4}}x\right)e^{-\sqrt{6|E|}\,\,x^2},
\end{equation}
where $H_n$ are Hermite polynomials. The wave functions
(\ref{cosmic-final-1}) are similar to the stationary quantum
wormholes as defined in \cite{Hawking}. However, neither of the
boundary conditions (\ref{boundary}) can be satisfied by these wave
functions.

For $\gamma=-2$ and $\alpha=0$ (dust regime), the WD equation
(\ref{sle2}) reduces to
\begin{equation}
\label{dust-2} -\psi'' + 144 \beta \psi =24Ea\psi,
\end{equation}
which can be written as
\begin{equation}
\psi'' -24 E \left(\frac{\displaystyle6 \beta }{\displaystyle
E}-a\right)\psi =0,
\end{equation}
by taking $x=\frac{\displaystyle6 \beta }{\displaystyle E}-a$ we have
\begin{equation}
\psi'' -24 E x\psi =0,
\end{equation}
which is again the Airy's differential equation \cite{Magnus}. Therefore, the physically acceptable
solution is
\begin{equation}
\psi(a)= \mbox{Ai}\left[(24E)^{1/3}\left(\frac{\displaystyle
6\beta}{\displaystyle E}-a\right)\right].
\end{equation}
We choose the first boundary condition (\ref{boundary}), which leads to
\begin{equation}
\mbox{Ai}\left[12\sqrt[3]{2}\beta E^{-2/3}\right]=0.
\end{equation}
Airy's function $\mbox{Ai}(x)$ has infinitely many negative zeros
$z_n$, therefore, the energy levels quantize and take the values of
\begin{equation}
E_n = \left(\frac{12\sqrt[3]{2}\beta}{z_n}\right)^{3/2},
\end{equation}
which exist only for negative values of $\beta$. The time-dependent
eigenfunctions take the form
\begin{equation}\label{dust-final-2}
\Psi_n(a,t)=e^{iE_n  t}
\mbox{Ai}\left[(24E_n)^{1/3}\left(\frac{\displaystyle
6\beta}{\displaystyle E_n}-a\right)\right].
\end{equation}
These solutions (\ref{dust-final-2}) show a quantum behavior for
small $a$ and a classical behavior for large $a$.

In $\gamma=-2$ and $\alpha=-1/3$ (cosmic strings) case, the WD
equation (\ref{sle2}) reduces to
\begin{equation}
\label{cosmic-2} -\psi'' + 144 \beta \psi =24Ea^{2}\psi.
\end{equation}
For $E<0$ the above equation can be written as
\begin{equation}
-\psi'' + 24|E|a^2\psi =-144 \beta \psi.
\end{equation}
This equation is identical to the time-independent Schr\"odinger equation for a harmonic oscillator with unit mass and energy $\lambda$, where
$2\lambda =-144 \beta $ and $\omega^2=24 |E|$. Therefore, the allowed values of $\lambda$ are $\omega(n+1/2)$ and the possible values of $E$ are
\begin{equation}
E_{n}=-\frac{216\beta^2}{(n+\frac{1}{2})^2}\,\, ,
\mbox{\hspace{0.8cm}} n=0,1,2,...\quad ,
\end{equation}
for $\beta<0$. Thus the stationary solutions are
\begin{equation}\label{cosmic-final-2}
{\Psi}_{n}(a,t)=e^{iE_{n}t}{\varphi}_{n}(a),
\end{equation}
\begin{equation}
{\varphi}_{n}(a)=H_n\left((24|E_n|)^{\frac{1}{4}}a\right)e^{-\sqrt{6|E_n|}\,\,a^2}.
\end{equation}
The solutions for odd $n$ satisfy the first boundary condition
(\ref{boundary}) and the appropriate wave packets can be constructed
by superposing these stationary solutions.

\section{Conclusion}
In this work we have investigated perfect fluid Stephani quantum
cosmological models. The use of Schutz's formalism allowed us to
obtain a Schr\"odinger-like WD equation in which the only remaining
matter degree of freedom plays the role of time. We have obtained
eigenfunctions and therefore constructed the acceptable wave packets
by appropriate linear combination of these eigenfunctions. The time
evolution of the expectation value of the scale factor has been
determined using the many-worlds and Bohmian interpretations of
quantum cosmology. We have shown that contrary to the classical
case, the expectation values of the scale factor avoid singularity
in the quantum case. At the end, we solved some interesting bound
state cases and found their discrete energy eigenvalues. We have
also shown that in some bound state cases, we may observe the
quantum effects in the large scales which correspond to the quantum
behavior at the late time cosmology.


\begin{thebibliography}{100}
\bibitem{1} G. Riess, et al, Astron. J. \textbf{116}, 1009 (1998).
\bibitem{2} A. Vilenkin, Phys. Rev. Lett. \textbf{53}, 1016 (1984).
\bibitem{3} R.L. Davies, Phys. Rev. D \textbf{36}, 997 (1997).
\bibitem{4} V. Silveira and I. Waga, Phys. Rev. D \textbf{50}, 4890 (1994).
\bibitem{5} M. Kamionkowski and N. Toumbas, Phys. Rev. Lett. \textbf{77}, 587 (1996).
\bibitem{6} R.R. Caldwell, D. Rahul and P.J. Steinhardt, Phys. Rev. Lett. \textbf{80}, 1582 (1998).
\bibitem{1-1} D. Kramer, H. Stephani, M.A.H. MacCallum, E. Herlt,Exact solutions of Einstein's field equations,(Cambridge University Press, Cambridge, U.K, 1980).
\bibitem{2-1} A. Krasi\'{n}ski, \textit{Inhomogeneous Cosmological Models}, (Cambridge University Press, Cambridge, U.K, 1998).
\bibitem{9} H. Stephani,  Commun. Math. Phys. \textbf{4}, 137 (1967).
\bibitem{3-3} A. Barnes, Gen. Rel. Gravit. \textbf{2}, 147 (1974).
\bibitem{12} A. Krasi\'{n}ski, Gen. Rel. Grav. \textbf{15}, 673 (1983).
\bibitem{14-2} V. G. Lapchinskii and V. A. Rubakov, Theor. Math. Phys. {\bf 33}, 1076 (1977).
\bibitem{R22-4} F. G. Alvarenga , J. C. Fabris , N. A. Lomes and G. A. Monerat , Gen. Rel. Grav. \textbf{34}, 651 (2002).
\bibitem{R22-6} A. B. Batista , J. C. Fabris , S. V. B. Gon\c{c}alves and J. Tossa, Phys. Rev. D \textbf{65}, 063519 (2002).
\bibitem{pedram} P.~Pedram, S.~Jalalzadeh and S.~S.~Gousheh, Phys. Lett. B, (2007) doi:10.1016/j.physletb.2007.08.077.
\bibitem{chap} P.~Pedram, S.~Jalalzadeh and S.~S.~Gousheh, Int.~J.~Theor.~Phys. DOI: 10.1007/s10773-007-9436-9.
\bibitem{11-2} B. F. Schutz, Phys. Rev. D {\bf  2}, 2762 (1970).
\bibitem{12-2} B. F. Schutz, Phys. Rev. D {\bf  4}, 3559 (1971).
\bibitem{R23-10} M. J. Gotay and J. Demaret, Phys. Rev. D \textbf{28}, 2402 (1983).
\bibitem{nivaldo-2} N. A. Lemos, J. Math. Phys. {\bf 37}, 1449 (1996).
\bibitem{R23-13} F. G. Alvarenga, A. B. Batista, J. C. Fabris and S. V. B. Gon\c{c}alves, Gen. Rel. Grav. \textbf{35}, 1659 (2003).
\bibitem{bohm1}P. R. Holland, \textit{The Quantum Theory of Motion: An Account of the de Broglie-Bohm Interpretation of Quantum Mechanics}, Cambridge University Press, Cambridge (1993).
\bibitem{bohm2}N. Pinto-Neto, Procedings of the VIII Brazilian School of Cosmology and Gravitation II, Edited by M. Novello (1999).
\bibitem{bohm3}A. Neumaier, \textit{Bohmian mechanics contradicts quantum mechanics}, [quantph/0001011].
\bibitem{bohm4}L. Marchildon, \textit{No contradictions between Bohmian and quantum mechanics}, [quantph/0007068].
\bibitem{bohm5}P. Ghose, \textit{On the incompatibility of quantum mechanics and the de Broglie-Bohm theory II}, [quant-ph/0103126].
\bibitem{7-2} R. Arnowitt, S. Deser and C. W. Misner, {\it Gravitation: An Introduction to Current Research}, edited by L. Witten, Wiley, New York (1962).
\bibitem{10} H. Stephani,  Commun. Math. Phys. \textbf{5}, 337 (1967).
\bibitem{14} M. P. D\c{a}browski, J. Math. Phys. \textbf{34}, 1447 (1993).
\bibitem{Godlowski} W. Godlowski, J. Stelmach, M. Szydlowski, Class. Quant. Grav. \textbf{21} 3953 (2004). astro-ph/0403534.
\bibitem{Stelmach}J. Stelmach, I. Jakacka, Class. Quantum Grav. \textbf{18}, 2643 (2001).
\bibitem{lemos1999} N. A. Lemos, F. G. Alvarenga, Gen. Rel. Grav. \textbf{31}, 1743 (1999), gr-qc/9906061.
\bibitem{lemos1998}F. G. Alvarenga, N. A. Lemos, Gen. Rel. Grav. \textbf{30}, 681 (1998).
\bibitem{18-2} G. A. Monerat, E. V. C. Silva, G. Oliveira-Neto, L. G. F. Filho, and N. A. Lemos, Phys. Rev. D \textbf{73}, 044022 (2006).
\bibitem{gradshteyn} I. S. Gradshteyn and I. M. Ryzhik, {\it Table of Integrals, Series and Products} (Academic, New York, 1980), formula 6.631-4.
\bibitem{many} F. J. Tipler, Phys. Rep. {\bf 137}, 231 (1986).
\bibitem{Magnus} W. Magnus, F. Oberhettinger and R. P. Soni, {\it Formulas and Theorems for the Special Functions of Mathematical Physics} (Springer, New York, 1966).
\bibitem{Coliteste} R. Coliteste, Jr., J. C. Fabris and N. Pinto-Neto,  Phys. Rev. D {\bf57}, 4707 (1998).
\bibitem{Hawking}S. W. Hawking and D. B. Page, Phys. Rev. D \textbf{42}, 2655 (1990).
\end{thebibliography}
\end{document}